\documentclass[12pt]{article}
\usepackage[latin1]{inputenc}
\usepackage{amsmath,amssymb,exscale}
\usepackage{psfrag,graphicx}
\input epsf

\def\be{\begin{equation}}
\def\ee{\end{equation}}
\def\bea{\begin{eqnarray}}
\def\eea{\end{eqnarray}}

\def\gappeq{\mathrel{\rlap {\raise.5ex\hbox{$>$}}
{\lower.5ex\hbox{$\sim$}}}}
\def\lappeq{\mathrel{\rlap{\raise.5ex\hbox{$<$}}
{\lower.5ex\hbox{$\sim$}}}}

\def\I{\rm 1\kern-.24em l}

\begin{document}
\topmargin -1.0cm
\oddsidemargin -0.5cm
\evensidemargin -0.5cm

\pagestyle{empty}
\begin{flushright}
May 2010
\end{flushright}
\vspace*{5mm}

\begin{center}
\vspace{3.cm}
{\Large\bf
Emergent~Gauge~Fields~in~Holographic~Superconductors
}
\vspace{2cm}\\
{\large Oriol Dom\`enech $^a$, Marc Montull $^a$, Alex Pomarol $^a$,}
\vspace{.4cm}\\
{\large  Alberto Salvio $^a$ and Pedro J. Silva $^{a,b}$}\\
\vspace{.8cm}
{\it {$^a$ Departament  de F\'isica and IFAE, Universitat Aut{\`o}noma de Barcelona, 08193~Bellaterra,~Barcelona}\\
{$^b$  Institut de Ci\`encies de l'Espai (CSIC) and Institut
d'Estudis Espacials de Catalunya (IEEC/CSIC),
Universitat Aut{\`o}noma de Barcelona,
08193 Bellaterra, Barcelona}}
\vspace{.4cm}
\end{center}

\vspace{1cm}
\begin{abstract}
Holographic superconductors
 have been studied so far in the absence of  dynamical electromagnetic fields, namely in the limit in which they coincide with holographic superfluids.
It is possible, however,  to
 introduce dynamical gauge fields if a Neumann-type  boundary condition is imposed on the AdS-boundary.
In   $3+1$ dimensions, the dual theory is a    $2+1$  dimensional CFT  whose spectrum contains a massless gauge field, signaling  the emergence of a  gauge symmetry.
We  study  the impact of a dynamical gauge  field  in  vortex configurations
where it is known to  significantly affect   the energetics  and phase transitions.
We calculate the critical magnetic fields $H_{c1}$ and $H_{c2}$, obtaining that holographic superconductors are   of Type~II ($H_{c1}<H_{c2}$).
We extend the study to $4+1$ dimensions where the   gauge field does not appear as an emergent phenomenon, but can be introduced, by a proper renormalization,  as an external dynamical field.
We also compare our predictions with those arising from a Ginzburg-Landau theory and identify the generic properties of  Abrikosov vortices  in holographic models.

\end{abstract}

\vfill
\eject
\pagestyle{empty}
\setcounter{page}{1}
\setcounter{footnote}{0}
\pagestyle{plain}

%
%
\tableofcontents
\section{Introduction}

The AdS/CFT correspondence  has  become a  powerful tool to study
strongly-coupled systems in different environments. Very recently
its applicability  has been also extended to
condensed matter systems.
In Ref.~\cite{Hartnoll:2008vx} a gravitational description of a superconductor was proposed
whose properties have been extensively studied in the last couple of years \cite{Hartnoll:2009sz}.
In these models a gauge U(1) symmetry is broken by a scalar field that turns on near the black-hole horizon.
This corresponds, in the dual description, to a global U(1) broken by
a scalar condensate.
Therefore, strictly  speaking, these  models describe either a superfluid \cite{Herzog:2008he} or
a superconductor in the gauge-less limit.

The reason for the absence of a dynamical gauge field in previously studied holographic superconductors is the chosen
AdS-boundary condition for the U(1) gauge field.
In most of  these  studies the gauge field  was chosen to be frozen  at the AdS-boundary
by imposing a Dirichlet boundary condition.
One can make, however, the gauge field  dynamical if one instead
imposes a Neumann-type  boundary condition at the AdS-boundary.
In this article we will make use of this option to
study  the role of  dynamical gauge fields in holographic superconductors
\footnote{
Neumann boundary conditions have been previously considered in holographic superconductors to study systems at fixed charge density. In these cases, however, the studied systems are homogeneous and therefore the  dynamical electric field vanishes.}.

In  a $3+1$ dimensional AdS space  it is known  that we can impose either a Dirichlet or a Neumann AdS-boundary condition   to quantize a gauge theory, being  both
related by an S-duality  \cite{Witten:2003ya}.
In the  Neumann case,  one finds a massless gauge field in the spectrum   that, by means of the AdS/CFT correspondence, can be considered to arise from a $2+1$ dimensional CFT.  It is therefore an emergent phenomenon.
In $4+1$ dimensions, however, a Neumann AdS-boundary condition for the
gauge  field is not well-defined since it leads to a non-finite Hamiltonian.
This will require, as we will show, to regularize the theory  and absorb  the divergencies in  local counterterms.
 In this case, the gauge field will not be an emergent phenomenon but just an external dynamical gauge field coupled to a $3+1$ CFT  \cite{ArkaniHamed:2000ds}.

An alternative way to understand the distinction between a gauge field
in a $2+1$ and $3+1$ CFT is to look at  the zero mode of the   Kaluza-Klein
expansion of the gauge field in the holographic superconductor model at  temperatures $T$ bigger than the critical temperature $T_c$.
For a stationary vector potential $a_i$ we find, after integrating over the extra dimension,
a  kinetic term given by
$\int d^dx {\cal F}_{ij}^2/(4 e_0^2)$,
where ${\cal F}_{ij}=\partial_ia_j-\partial_ja_i$ and
\be
\frac{1}{e^2_0}=\frac{3}{4 \pi g^2T}\ \ \ \ \ \text{for  $d=2+1$}\ ,\ \ \ \
\frac{1}{e^2_0}=-\frac{L}{g^2}\ln\left.(z \pi T)\right|_{z=0}\ \ \ \ \ \text{for  $d=3+1$}\, .
\label{running}
\ee
Here $g$ is the gauge coupling in the AdS model and $L$ is the AdS radius.
In $2+1$ dimensions
this massless gauge boson mode has a finite  norm and therefore remains in the spectrum, while  for $d=3+1$
this  is a non-normalizable mode and disappears from the set of dynamical degrees of freedom.
To keep this mode in  $3+1$ dimensions we must then make its norm finite, for example by adding  local counterterms.

 The impact of a dynamical gauge field  in  superconductors is expected to be important in inhomogeneous  configurations. For this reason   we will concentrate  here on  the vortex configurations of the holographic models.
We will explicitly analyze the cases $d=2+1$ and $d=3+1$, introducing, when studying the  $d=3+1$ holographic superconductor, local counterterms to render the norm of $a_i$ finite. This will allow us to  explicitly see that the dynamical magnetic field $B$ plays an important role in reproducing some of the known features of superconductor vortices, such as the exponential damping  of $B$ inside the superconductor.
  We will  also show that  the dynamical gauge field  changes the vortex configurations
of the holographic models, making  the energy and, correspondingly, the first critical magnetic field
$H_{c1}$ independent of the  sample size, as expected from a true Abrikosov vortex.
The properties of the new configurations  are qualitatively similar to those arising from a
Ginzburg-Landau (GL) theory  \cite{Ginzburg:1950sr}, although we find important quantitative differences in the size of the vortex core,  the profile of $B$  flowing through the vortex and  $H_{c1}$.

The effect of an external magnetic field on holographic superconductors has been considered in  \cite{Hartnoll:2008kx,Albash:2008eh}
and vortex solutions have been studied before for $d=2+1$  in Refs.~\cite{Montull:2009fe, Keranen:2009re}.
In all these cases there was  no dynamical
electromagnetic (EM) field, and  therefore the vortices were not true Abrikosov configurations but just superfluid vortices.
Only Ref.~\cite{Albash:2009iq} showed, for $d=2+1$, a non trivial profile for the magnetic field of the form of a vortex magnetic tube; it is however unclear the  origin of this magnetic field
and its relation with Abrikosov configurations.

The organization of the paper is as follows.
In Section~2 we  describe, similarly  in spirit to Ref.~\cite{Weinberg:1986cq},
 effective field theories
 from which we can obtain
 model-independent  properties of  superconductors    and superfluids,
 and their vortex configurations.
 This will also help us to make contact with the  GL predictions
 \footnote{What we mean with GL theory
in the case of superfluids is the limit of  frozen magnetic fields in the GL theory for superconductors.}.
In Section~3 we present the holographic model
and  explain how to introduce dynamical gauge fields.
We then  focus on the  holographic superfluid and superconductor (Abrikosov) vortex in   $d=2+1$ and $d=3+1$ dimensions, comparing them
 with those   of  the GL theory.
We  calculate the energy of these configurations to
  find the critical magnetic fields $H_{c1}$ and $H_{c2}$
and show that the holographic superconductors are always of Type II.
In Section~4 we present a summary of the results and other concluding remarks.


\section{Effective theories of superfluids and superconductors}\label{ET}

We are interested in the effective theory for time-independent
configurations of a  U(1) gauge field $a_{\mu}=(a_0,a_i)$, where $i,j=1,...,
d-1$, and
 a scalar  field  $\Phi_{\rm cl}$ whose non-zero  value will be responsible for the U(1) breaking.
The effective action at finite temperature $T$ for $a_i$ and  the order parameter
$\Phi_{\rm cl}$, obtained after integrating out all the other fields of
the theory, depends  on an effective  Lagrange density constructed
from gauge-invariant operators:
\begin{equation}
\Gamma= \beta\int d^{d-1}x \,{\cal
L}_{\rm eff}\ ,\ \ \ \
{\cal L}_{\rm eff}= {\cal L}_{\rm eff}\big({\cal F}^2_{ij}, |D_i\Phi_{\rm
cl}|^2, |\Phi_{\rm cl}|, ...\big)\, , \label{genfun}
\end{equation}
where
 $D_j=\partial_j-ia_j$ and  $\beta=1/T$. Eq.~(\ref{genfun}) is defined in some
renormalization scheme. We will be assuming that this theory depends
only on two mass-scales, $\mu$ (that later we will  associate with a
chemical potential) and the temperature $T$.

The generic effective theory given by Eq.~(\ref{genfun}) simplifies  in two
limits. In the limit of small fields (as compared to $\mu$ and $T$),
this theory approximates to the GL theory
\begin{equation}
\Gamma_{\rm GL}=\beta \int d^{d-1}x \Big\{\frac{1}{4e^2_0} {\cal F}^2_{ij}+|D_i\Phi_{\rm GL}|^2   +V_{\rm GL}(|\Phi_{\rm GL}|)\Big\}\, .
\label{gl}
\end{equation}
The GL field $\Phi_{\rm GL}$ is defined  to be canonically normalized, $\Phi_{\rm GL}= \sqrt{h_0}\,\Phi_{\rm cl} $, where $h_0$ is a positive constant,
 and \begin{equation} V_{\rm GL}=-\frac{1}{2\xi_{\rm GL}^2} |\Phi_{\rm GL}|^2+b_{\rm GL}|\Phi_{\rm GL}|^4.\end{equation}
This approximation becomes reliable, for example,  close to the
critical temperature $T\lesssim T_c(\mu)$ where the ``condensate"
$\Phi_{\rm cl}$ has a small value. The other useful  limit
corresponds to slowly varying fields, which implies  that
$D_i\Phi_{\rm cl}$ and   ${\cal F}_{ij}$ are small and
   \begin{equation}
\Gamma\simeq \beta \int d^{d-1}x\, h(|\Phi_{\rm
cl}|)\Big\{\frac{1}{4e^2(|\Phi_{\rm cl}|)} {\cal F}^2_{ij}+|D_i\Phi_{\rm
cl}|^2   +W(|\Phi_{\rm cl}|)\Big\}\, , \label{limit}
\end{equation}
where $h$, $W$ and $e$ are  generic functions of  $|\Phi_{\rm cl}|^2$. In the limit of small fields,
we obtain the GL theory:  $h(|\Phi_{\rm cl}|)\rightarrow h(0)= h_0$,
$W(|\Phi_{\rm cl}|)\rightarrow V_{\rm GL}(|\Phi_{\rm GL}|)/h_0$ and
$e^2(|\Phi_{\rm cl}|)\rightarrow e^2(0)= h_0 e^2_0$.
When Eq. (\ref{gl}) and/or Eq. (\ref{limit})
are applicable they can be used to extract model independent features of superconductors and superfluids.

Consider first the case where we are at large temperatures $T>T_{c}(\mu)$; here the condensate $\Phi_{\rm cl}$ is zero,  corresponding to the  ``normal" phase.
By decreasing the temperature, $T<T_{c}(\mu)$ at zero magnetic field, the modulus of the scalar field $\psi_{\rm cl}=|\Phi_{\rm cl}|$ will get a nonzero constant value $\psi_\infty$.
For this homogeneous configuration the effective action in Eq.~(\ref{limit}) is obviously a good approximation of the theory \footnote{Notice however that,  generically, this is not the case for the GL theory in Eq.~(\ref{gl}).}, and  the  value of  $\psi_\infty$ is determined by
 the minimum of the potential $V= h W$:
\begin{equation}
\frac{\partial V}{\partial \psi_{\rm cl}}(\psi_\infty)=0\ .\ \
\label{minimum}
\end{equation}
This configuration corresponds to   the superfluid/superconductor phase.
Two important parameters describing these systems are  $\xi$ and $\lambda$,  defined as
\begin{equation}
\frac{1}{\xi^2} = \frac{1}{2 h(\psi_{\infty})} \frac{\partial^2  V}{\partial \psi^2_{\rm cl}}(\psi_\infty) >0\,, \quad \lambda=\frac{1}{\sqrt{2}e(\psi_\infty)\, \psi_\infty}\,.
\label{lambda}
\end{equation}
These quantities exactly correspond to the inverse mass of the scalar $\psi_{\rm cl}$ and $a_i$ respectively.

 In this work we will be considering time-independent vortex configurations with  cylindrical symmetry as the main example of our theoretical framework.
We define $(r,\phi)$ as the polar coordinates restricted to $0\leq r\leq R$, $0\leq \phi<2\pi$. We will always consider the case $\xi \ll R$ and, in the superconductor case, also $\lambda \ll R$.
We take the  Ansatz
\begin{equation}
a_\phi=a_\phi(r) \ , \ \  \Phi_{\rm cl}=e^{in\phi}\psi_{\rm cl}(r)\,  ,\label{ansatzEffective}
\end{equation}
where $n$ is an integer, and all other gauge components are set to zero. For $n\neq 0$, the fields $a_\phi(r)$ and $\psi_{\rm cl}(r)$,
satisfying the equations of motion from the Lagrangian in (\ref{genfun}),
describe a straight vortex line centered at $r=0$. If we insert the Ansatz (\ref{ansatzEffective}) into the action in (\ref{limit}) we obtain
\begin{equation}
\Gamma \simeq 2\pi V^{d-3}\beta\int^R_0 dr r\, h(\psi_{\rm cl})\Big\{\frac{1}{2e^2(\psi_{\rm cl})r^2}(\partial_r a_\phi)^2+(\partial_r\psi_{\rm cl})^2+\frac{1}{r^2}(n-a_\phi)^2\psi_{\rm cl}^2   +W(\psi_{\rm cl})\Big\}\, ,
\label{limitp}
\end{equation}
where $V^{d-3}$ is the volume of the space orthogonal to the plane
$(r,\phi)$. Here the current is given
by~\footnote{Notice that we have defined the current to include the
kinetic term of the gauge field. This is done in order to facilitate
our treatment for both, dynamical and non-dynamical gauge fields.}
\begin{equation}
 J_\phi=-\frac{1}{\beta}\frac{\delta{\Gamma}}{\delta  a^{\phi}}=2h(\psi_{\rm cl})(n-a_\phi)\psi_{\rm cl}^2+
r \partial_r \left(\frac{h(\psi_{\rm cl})}{e^2(\psi_{\rm cl}) r} \partial_ra_{\phi}\right)\,. \label{currentDef}
\end{equation}
In the vortex case ($n\not=0$) $\psi_{\rm cl}$ goes to $\psi_{\infty}$ far away from the vortex center; this corresponds to the physical fact that a vortex line destroys superfluidity/superconductivity
only in a region close to its center.   The details of the vortex configurations depend on whether the field
$a_\phi$ is dynamical as in the superconductor case, or just a non-dynamical background as  in a superfluid system. We  consider the two cases in turn.


\subsection{Superfluid vortex}
For superfluids  the modulus and phase of $\Phi_{\rm cl}$ are respectively associated with
the density $n_s$ and velocity $v_i$ of the superfluid. In the limit of slowly varying fields, Eq.~(\ref{limit}), we define them
as
\footnote{Right dimensions are obtained by putting appropriate
powers  of the ``boson" mass causing the superfluidity.}
\begin{equation}
n_s(|\Phi_{\rm cl}|)=2|\Phi_{\rm cl}|^2h(|\Phi_{\rm cl}|)\  ,\ \
v_i=\partial_i {\rm Arg}[\Phi_{\rm cl}]\, .
\label{defns}
\end{equation}
The field $a_\phi$ is not dynamical; it just represents an external
angular velocity performed on the superfluid.
This is implemented by working in a rotating frame with a constant  angular velocity
$\Omega=a_\phi/r^2$.
In going from the static to the rotating frame the
angular velocity of the superfluid is changed accordingly: $v_\phi\rightarrow v_\phi-\Omega r^2$.  The current is then given by $J_\phi=n_s(v_\phi-\Omega r^2)$.

Superfluid dynamics coincides with those of a superconductor in the limit in which
the EM field is frozen to certain values.
This is achieved by  taking the limit $e\rightarrow0$
 while keeping
the external magnetic field $B=\partial_r a_\phi/r$ constant.
In this limit the correspondence between the superfluid and the superconductor systems is given by
\begin{equation}
\Omega\leftrightarrow B/2\ , \ \    L_\bot\leftrightarrow 2M\, ,
\label{dict}
\end{equation}
where $L_\bot$ is the angular momentum and $M$ the magnetization  of the system
in the direction perpendicular to the  $(r,\phi)$ plane.
In the rest of this section we will use the superconductor notation.

Vortices  correspond to configurations with $n\not=0$ where  $\psi_{\rm cl}$
varies from zero (at $r=0$)  to $\psi_\infty$ at large $r$. The exact solution
depends on the specific effective action and therefore it is very model dependent.
Nevertheless, we can obtain the behavior of $\psi_{\rm cl}$ in the limit $r\rightarrow 0$
and  for large $r$.
Indeed, for $r\rightarrow 0$, the condensate goes to zero and the GL action can be applied
to obtain
\begin{equation}
\psi_{\rm cl}\propto r^{|n|}\, . \label{rnbehaviour}
\end{equation}
For large $r$, we can use Eq.~(\ref{limitp})
to obtain, in the absence of rotation ($B=0$),
\begin{equation}
\psi_{\rm cl}\simeq\psi_\infty\left[1-n^2\frac{\xi^2}{r^2}\left(1+\frac{\psi_\infty}{2h(\psi_\infty)} \, \frac{\partial h}{\partial \psi_{\rm cl}}(\psi_\infty)
\right)\right]\, , \label{xibehaviour}
\end{equation}
showing that $\sim\xi$  gives the  size  of the vortex core radius.
For  $B=0$ the free energy  per unit of volume $V^{d-3}$, $F_n$,   is dominated by the
third  term of Eq.~(\ref{limitp}):
\begin{equation}
F_n-F_0\simeq 2\pi \int_0^R \frac{dr}{r} h(\psi_{\rm cl})\, n^2\psi_{\rm cl}^2\simeq \pi n_s(\psi_\infty)\, n^2\int_\xi^R \frac{dr}{r}= \pi n_s(\psi_\infty)\, n^2 \ln \left(R/\xi\right)\, ,
\label{logdiv}
\end{equation}
 that depends logarithmically on the size of the superfluid sample $R$.
  This shows that superfluid vortices are not finite-energy configurations in the limit $R\rightarrow \infty$.
For  $B\not=0$, we have to consider the free energy as a function of the angular velocity, obtaining \footnote{In the superfluid case, this is the correct expression for the energy calculated in the co-rotating system with respect to the container.}
\begin{equation}
F_n(B)=F_n(0)-\int^B_0 M_n(B) dB\, ,
\label{newform}
\end{equation}
where $M_n(B)$ is the magnetization  (angular momentum from Eq.~(\ref{dict})) of the $n$-vortex configuration:
\begin{equation}
M_n=\pi\int dr\, r  J_\phi\, .
\label{mn}
\end{equation}
The value of $M_n$   is approximately  given by

\begin{equation}
M_n\simeq \pi n_s(\psi_\infty)\int_\xi^Rdrr\left(n-\frac{r^2}{2}B\right) \simeq \pi n_s(\psi_\infty) \left(\frac{nR^2}{2}-\frac{R^4}{8}B\right)\, ,
\label{mn2}
\end{equation}
that leads to
\begin{equation}
F_n(B)\simeq F_0(B)+\pi n_s(\psi_\infty)\left( n^2 \ln \left(R/\xi\right)-\frac{1}{2}   nR^2B\right)\, .
\label{Fn2}
\end{equation}
From this formula we can easily  calculate the critical angular velocity  $B_{c1}$
above which the vortex configuration is energetically favorable.
This is given by the $B$ field at which $F_1=F_0$:
\begin{equation}
B_{c1}\simeq\frac{2}{R^2}\ln\left( R/\xi\right)\, .
\label{bc1}
\end{equation}
We observe that $B_{c1}\rightarrow 0$ when the size of the sample goes to infinity, that is $R\rightarrow \infty$.

By increasing $B$, more and more  vortices are formed up to a critical value $B_{c2}$
at which the normal phase is favorable. In the limit $B\rightarrow B_{c2}$ the condensate goes to zero, and
the GL theory can be applied. One obtains, with a standard textbook derivation,
\begin{equation}
B_{c2}=\frac{1}{2\xi_{\rm GL}^2} \, \label{xiBc2}\, .
\end{equation}

\subsection{Superconductor vortex}\label{superconductor vortex}

For superconductors,  $a_i$, and correspondingly the magnetic field $B$, are dynamical fields \footnote{In this case we call  the dynamical magnetic field $B$,
while we keep $H$ for the external magnetic field.}.
Superconductor vortex configurations are therefore described by two fields, $\psi_{\rm cl}$ and $a_\phi$.
The value of $a_\phi$ varies from zero (at $r=0$)  to $n$ at infinity, canceling the logarithmic divergence in Eq.~(\ref{logdiv}) and  making the vortex energy
 finite  in the limit  $R\rightarrow \infty$.

Like in the superfluid case, we can obtain the behavior of the fields at small and large $r$ in a model independent way.
At small $r$, the condensate drops to zero and the GL action  can be used; in this limit one can derive

\be
\psi_{\rm cl}\propto r^{|n|}\ ,\ \ a_\phi\propto r^2\, .
\label{limitr0}
\ee
At large $r$ the situation is more complicated. If one uses Eq. (\ref{limitp}) it is possible to show that the fields have the following large $r$
behavior
\begin{equation}
\psi_{\rm cl}\simeq \psi_\infty+\frac{\psi_1}{\sqrt{r}}e^{-r/\xi'}\ , \ \ \  a_\phi\simeq n+a_1\sqrt{r}e^{-r/\lambda'}\, ,
\label{limitrinfty}
\end{equation}
with $\xi'=\xi$,  $\lambda'=\lambda$
and $\psi_1$ and $a_1$ being  constants.
Nevertheless, Eq.~(\ref{limitrinfty}) shows that
higher derivatives are not negligible with respect to the first and second derivatives that  we included in Eq. (\ref{limitp}). Indeed,   we have
\be
\frac{\partial_r^n a_\phi}{\mu^n}  \sim \frac{1}{(\lambda \mu)^{n-1}}
\frac{\partial_r a_\phi}{\mu}\sim
\frac{\partial_r a_\phi}{\mu}\, ,
\ee
where we have assumed, based on dimensional grounds, that the scale $\mu$ suppresses the higher-dimensional operators, and that  $\lambda$ is of order $1/\mu$. A similar situation happens  for $\psi_{\rm cl}$.
We are therefore led to the conclusion that we cannot neglect
higher-derivative terms  to describe the large $r$ behavior of the fields.
 Including them, the equations of motion can (formally)  be written as
\be {\cal M}(\Box)\psi_{\rm cl} \simeq \frac{1}{\xi^2} \left(\psi_{\rm cl}- \psi_{\infty}\right) \ , \ \ \
{\cal N}(\Box)a_{\phi} \simeq \frac{1}{\lambda^2} \left( a_{\phi}-n \right)\, , \label{hdEOM}
\ee
where ${\cal M}$ and ${\cal N}$ are unknown functions and the box operator acts on  $\psi_{\rm cl}$ and $a_\phi$ as
\be
\Box\psi_{\rm cl}=\frac{1}{r}\partial_r \left( r \partial_r\psi_{\rm cl} \right)\ , \ \ \
\Box a_{\phi} = r\partial_r\left(\frac{1}{r} \partial_r a_{\phi}\right)\, .
\ee
Fortunately,   the solutions to the equations above are also of the form of Eq.~(\ref{limitrinfty}) but with   $\lambda'$ and $\xi'$  generically different from $\lambda$ and  $\xi$.
In other words,  the effect of the higher-derivative terms  is just to change the values of  $\lambda'$ and $\xi'$.
 From this large $r$ behavior we can see that the
radius size of the vortex core
and the radius size of the magnetic tube passing through the vortex (the penetration length)  are, respectively, characterized  by $\xi'$ and  $\lambda'$.

To calculate the external magnetic field $H$ at which the vortex
configuration is energetically favorable we must obtain the Gibbs
free energy. This is given in terms of the free energy $F$ by
\begin{equation}
G[J_{ext}]=F - \int d^{d-1} x\,  a_iJ_{ext}^i \, ,
\label{Hext}
\end{equation}
where  $J_{ext}^i$ is an  external current  coupled to the gauge field $a_i$.
We can relate  $J_{ext}$ to  the external magnetic field $\vec H$ that it produces, through
\footnote{In this Maxwell equation of the external field we use  $e_0$,
defined as the electric charge in the normal phase ($\psi_{\rm cl}=0$),
to  guarantee that when  $T\rightarrow T_c$, the magnetic field $B$ approaches $H$.}
\be
\nabla\times \vec H=e_0^2 J_{ext}\, .
\label{maxH}
\ee
 Then we end up with the following Gibbs free energy per unit of volume $V^{d-3}$ of the vortex configuration
\begin{equation}
G_n[H]= F_n-\frac{1}{e_0^2}\int rdrd\phi\, B H=F_n-\frac{2\pi n}{e_0^2}H\, ,
\label{gibbs}
\end{equation}
where we have  used the magnetic flux condition
$\int rdrd\phi B=2\pi n$ and assumed that
  $H$ is constant.
The critical $H_{c1}$ is defined as the value of $H$ at which $G_1=G_0$ that
corresponds to
\be H_{c1}=
\frac{e_0^2}{2\pi} (F_1-F_0)\, . \label{hhc1} \ee
The exact value of $F_1-F_0$ depends strongly  on the model and therefore
$H_{c1}$ can only be calculated once the  model is specified.

The minimum value of $H$ for which the energetically
favorable phase is the normal phase is also, as in the superfluid
case,  $H_{c2}=1/(2\xi^2_{\rm GL})$.
The superconductors that have
energetically favorable vortex solutions, that is $H_{c1}< H_{c2}$,
are  called Type II superconductors, while the others are
 called Type I. When the external field is
slightly smaller than $H_{c2}$ the condensate has a small value and
the GL theory can be applied to predict that Type II superconductors
present a triangular lattice of vortices \cite{triangular}.
Superfluids can be considered as deep Type II superconductors and
therefore they also present a triangular lattice of vortices.
 We will show that
holographic superconductors  are of Type II.


\section{Holographic superfluids and superconductors}

The holographic theory that we want to study is defined
\cite{Hartnoll:2008vx,Horowitz:2008bn} by  a charged scalar  $\Psi$
coupled to a U(1) gauge field $A_{\alpha}$  in $d+1$ dimensions ($\alpha, \beta
=0,1,..., d$) and an action
given by \be S=\int d^{d+1}x\, \sqrt{-G}\left\{{{1\over16\pi
G_N}}\left(R-\Lambda\right)\,+\frac{1}{g^2}\mathcal{L}\right\}\ ,\
\ \hbox{with }\ \ \mathcal{L}=-{1\over4}{\cal F}_{\alpha \beta}^2-\frac{1}{L^2}|D_\alpha\Psi|^2\, . \label{action} \ee $G_N$ is
the gravitational Newton constant and  the cosmological constant
$\Lambda$ defines the asymptotic AdS radius $L$ via the relation
$\Lambda=-{d(d-1)/L^2}$; moreover we introduced ${\cal F}_{\alpha \beta}= \partial_\alpha A_\beta-\partial_\beta A_\alpha$  and $D_{\alpha}=\partial_{\alpha}-iA_{\alpha}$.  For simplicity,    we have not added any
potential for the scalar. We will later discuss the implications of
including these terms. We will work in the limit  $G_N\rightarrow0$
and $g\rightarrow 0$ taken such that  the gravitational effect of
${\cal L}/g^2$ can be neglected. In this limit the metric is given
by an AdS-Schwarzschild black hole (BH):
\begin{equation}
 ds^2= \frac{L^2}{z^2}\left[-f(z)dt^2+dy^2 \right]+\frac{L^2}{z^2f(z)}dz^2 \,,\ \ \ f(z)=
 1-\left(\frac{z}{z_h}\right)^d \, ,\label{AdSBH}
 \end{equation}
where $t$ is time, $z$ is the holographic direction such that the
AdS-boundary  occurs at $z=0$, while the BH horizon is at $z=z_h$
and $dy^2$ stands for the $d-1$ dimensional flat metric. Since we
are interested in the  theory at finite temperature, we will
perform the Euclidean continuation with compact time $it \in  [0,
1/T ]$ where $T=d/(4 \pi z_h)$.

\subsection{The AdS/CFT correspondence and dynamical gauge fields}

This $d+1$ dimensional theory has  a dual interpretation in terms of a $d$ dimensional
CFT at nonzero temperature. The AdS/CFT
dictionary  relates the properties of the AdS gravitational
theory with those of the CFT. In particular,
 the fields $A_\mu$ and $\Psi$ evaluated on
the AdS-boundary correspond to fields external to the CFT:
 \be
a_\mu=A_\mu|_{z=0}\ , \ \  s=\Psi|_{z=0}\, .
\label{bcz0} \ee
They are coupled to CFT operators through the
interaction terms  $a_\mu \hat J^\mu+s {\cal O}$.
The operator $\hat J_\mu$ corresponds to the U(1) current of the CFT theory, while ${\cal
O}$ is a  CFT operator charged under the U(1) with
Dim$[\mathcal{O}]=3(4)$ for $d=3(4)$. Having chosen a nonzero mass
for the scalar $\Psi$ in Eq.~(\ref{action}), would  have
corresponded to take another dimensionality for ${\cal O}$. We do not expect however any   important qualitative difference for other choices of the mass.
The dual
CFT theory, if it exists,  is supposed to be  strongly coupled and
the  limit $g\rightarrow 0$ in the AdS theory corresponds to  be
working at the planar level in the CFT.

Integrating over the CFT fields, one can obtain the free energy $F[a_\mu,s]$
from which the  vacuum expectation values (VEV) of the CFT operators can be extracted.
In the gravity  side, $F[a_\mu,s]$   is  obtained from
the  $d+1$ dimensional AdS Euclidean action $S_E[a_\mu,s]$ evaluated
with all bulk fields on-shell restricted to Eq.~(\ref{bcz0}):
\begin{equation}
F[a_\mu,s]= T\, S_E[a_\mu,s]\, ,
\label{freee}
\end{equation}
from which we  obtain the VEVs of the currents
\be
\langle \hat J_\mu\rangle=\frac{L^{d-3}}{g^2}z^{3-d}{\cal F}_{z\mu}|_{z=0}\ , \ \  \langle{\cal
O}\rangle=\frac{L^{d-3}}{g^2}z^{1-d}D_z \Psi^*|_{z=0}\, .
\label{operator} \ee
The matching with the effective theory of Section~2 is
straightforward: the gauge field $a_i$ of Eq.~(\ref{bcz0}) is
identified with that in Eq.~(\ref{genfun}), while $\langle \hat
J_i\rangle$ and $\langle{\cal O}\rangle$ of Eq.~(\ref{operator}) are
identified respectively with $-\beta^{-1}\delta \Gamma/\delta a^i$ and $\Phi_{\rm cl}$ when renormalized in the
same scheme.

In the AdS/CFT correspondence, the external fields in   Eq.~(\ref{freee}) are  considered to be frozen
background fields.  This is suited for holographic  superfluids, but not for
superconductors that require the presence of dynamical gauge fields coupled to the CFT.
It is easy however to promote the external $a_\mu$ field to a  dynamical field.
This corresponds to integrating over it in the path integral:
\be
G[s,J_{ext}]=-T\ln\int Da\,  e^{-\beta F[a_\mu,s]+\int d^{d}x \left[-\frac{1}{4e_b^2}{\cal F}^2_{\mu \nu}+a_\mu J^\mu_{ext}\right]}\, ,
\label{newg}
\ee
where, for generality, we have added to   $F[a_\mu,s]$ a ``bare" kinetic term for $a_\mu$
  ($e_b$ denotes
the bare electric charge), and have coupled it to a background external current
$J^\mu_{ext}$ to define a Gibbs energy. Working in  the semiclassical  approximation \footnote{The semiclassical approximation is valid in the  limit $g\rightarrow 0$ and
 $e_b\rightarrow 0$.}, Eq.~(\ref{newg}) leads to the  Maxwell equation for the gauge field $a_\mu$:
\be
\langle \hat J^\mu\rangle+\frac{1}{e_b^2}\partial_\nu {\cal F}^{\nu \mu}+J^\mu_{ext}=0\, ,
\label{maxwell}
\ee
where we have used that  $\langle \hat J^\mu\rangle=-\delta F/\delta a_\mu$.
Let us see how the above procedure  can be implemented  in the gravity  side.
Using  Eq.~(\ref{operator}), we can  write Eq.~(\ref{maxwell})  as the following AdS-boundary  condition:
\be
\frac{L^{d-3}}{g^2}z^{3-d}{\cal F}_z^{\,\,\, \mu} \Big|_{z=0} +\frac{1}{e_b^2}\partial_\nu {\cal F}^{\nu \mu}\Big|_{z=0}+J^\mu_{ext}=0\, .
\label{maxwell2}
\ee
This is a  boundary condition of  Neumann type that, in order to be consistent  with the variational principle,  requires the AdS model
to include the following extra terms  on the AdS-boundary:
\be
\int d^{d} x \left[
-\frac{1}{4e_b^2}{\cal F}_{\mu\nu}^2+A_\mu J_{ext}^\mu\right]_{z=0}\, .
\label{extrat}
\ee
Therefore  the Gibbs free energy
is given by
 the AdS Euclidean action $S_E$ including the additional terms Eq.~(\ref{extrat})
 evaluated on-shell with the bulk fields  restricted  to the AdS-boundary condition Eq.~(\ref{maxwell2}).

In the particular case of $d=2+1$, and in the limit where
 $e_b/g\rightarrow \infty$ (not adding a  kinetic term for the gauge field on the AdS-boundary),
 one can show that the theory defined by Eq.~(\ref{newg}) preserves   conformal symmetry.
In this case
the original CFT and the $a_\mu$   can be considered as part of a new CFT.
Another way to understand this result is given in Ref.~\cite{Witten:2003ya}.
There  it was shown that
there are two ways to quantize a gauge field in the four dimensional
AdS. We can either impose a Dirichlet or a Neumann boundary condition at $z=0$.
Each option is associated with a different CFT,  S-dual to each other,
with different global U(1). While in  the first option (Dirichlet boundary condition) the gauge field $a_\mu$  is
a background field,
in the second one   (Neumann boundary condition) the gauge field is truly  dynamical \cite{Witten:2003ya}.
In this latter case   the gauge field  arises   from the  CFT as a composite
state,   as shows the fact that its kinetic term is induced  by the AdS bulk dynamics. In other words, this local U(1) appears as  an emerging phenomenon.
 As emphasized in Ref.~\cite{Witten:2003ya},  this CFT, which includes a dynamical gauge field, has also
 a global U(1)  with an associated conserved current given  by $\tilde J^\mu=\epsilon^{\mu\nu\sigma}\partial_\nu a_\sigma$, and should not be confused with the emerging local  U(1).
Here we also observe that  the emergence of the dynamical U(1) can be understood without using conformal invariance. Even for a warped space different from AdS, the
massless zero-mode of a gauge field  in  $3+1$ dimensions
has finite norm, corresponding then to a composite state in the dual $d=2+1$ theory.
This is  related to the fact that  the gauge interaction in  $d=2+1$ is a relevant operator
and therefore is  dominated by  IR physics.
It is thus possible to send $e_b/g$ to infinity in this case.

For $d=3+1$, the situation is different.
The current $\langle \hat J_\mu\rangle$ contains a logarithmically divergent piece given by
(in the gauge $A_z=0$)
\be
\frac{1}{z} \partial_z A_\mu\Big|_{z=0} = -\partial^{\nu}{\cal F}_{\nu \mu} \ln z\Big|_{z=0}   +
 ... \,\label{behaviourAphi4}\, .
\ee
The appearance of  the  logarithmic divergence
was already expected from the calculation of the kinetic term of $a_i$ in Eq.~(\ref{running}).
This can also  be   understood by looking at the dual CFT interpretation of the gravitational
theory. Indeed, at short distances (smaller than $1/T$) this dual theory is a $3+1$ dimensional relativistic  theory  charged under a U(1).
 At the quantum level an external  $a_\mu$ gauging this U(1) receives  corrections to its self-energy  that in momentum space go as
\be
\Pi(p^2)\simeq p^2\ln \left(p^2/\Lambda_b^2\right)\, ,
\ee
where $p^2$ is the 4-dimensional momentum of the gauge field and  $\Lambda_b$ is a momentum cut-off that regularizes a logarithmic divergence. Therefore $a_\mu$ is a state
of infinite norm.
If our intention is to keep the external gauge field in the theory we must renormalize it.
A possible way to do so
 is to place a UV-brane at finite $z>0$ as in Randall-Sundrum models
\cite{Randall:1999ee}. Alternatively,  we can  absorb  the divergence in the local  counterterm
of Eq.~(\ref{extrat}), i.e.,  defining  the bare coupling $e_b$ as
\be
\frac{1}{e_b^2}=\frac{1}{e_0^2}+\frac{L}{g^2} \ln z|_{z=0}+finite\ terms\, ,
\ee
where    $e_0$ denotes  here and thereafter  our renormalized  (physical) electric charge at the normal phase ($\psi_{\rm cl}=0$).
Contrary to the $d=2+1$ case,  the presence of the gauge field $a_\mu$
breaks conformal invariance; therefore
 the  gauge field cannot be considered an emerging phenomenon
but just a new external state coupled to the CFT \cite{ArkaniHamed:2000ds}.
 The same  is   true for any $d>4$.

We are now ready to study  models  of holographic  superconductors.
We are interested in vortex configurations where, as we said, the effects of
dynamical gauge fields are important.
We will however  present first the holographic superfluid
vortex configurations for both $d=2+1$ and $d=3+1$.
This will be useful to clarify previous results in the literature \cite{Montull:2009fe}, showing
that  these holographic vortices  fulfill the  expectations of
Section~2 for  configurations without dynamical gauge fields.
Then, we will present our   main result: the Abrikosov  superconducting vortex.

\subsection{The vortex Ansatz}

For  both the superfluid and the superconductor we will demand \be
s=0\ ,\ \ a_0=\mu\, ,\label{bcboth} \ee where $\mu$ is a constant.
We fix $s=0$  since we are only interested in the case in which the
U(1) symmetry is broken dynamically by the VEV of $ {\cal O}$. The
constant $\mu$ plays the role of a chemical potential.
As shown in Ref.~\cite{Hartnoll:2008vx,Horowitz:2008bn}, a nonzero
$\mu$ is necessary in order to induce, at temperatures smaller than
some critical temperature $T_c(\mu)$, a nonzero value for
$\langle{\cal O}\rangle$    and to have the system in a
superfluid/superconductor phase. This critical temperature is given
by \cite{Hartnoll:2008vx,Horowitz:2008bn}
\be T_c\simeq 0.03(0.05) \mu \quad \mbox{for}\,\,\, d=3(4)\, .
\label{T_c}\ee
Notice that $\mu \not= 0$ breaks  the conformal symmetry  of the system
and, together  with  $T$, set the  scales of the model.

To obtain vortex solutions we take the Ansatz \cite{Montull:2009fe,Albash:2009iq}
\be \Psi=\psi(z,r)e^{i n\phi}\  ,\quad A_0=A_0(z,r)\ , \quad
A_{\phi}=A_{\phi}(z,r)\, , \label{ansatz} \ee
and the other components of $A_{\alpha}$  set equal to zero. As in the
previous section, $n$ is an integer and a  vortex corresponds to
$n\neq 0$. We will be working in polar coordinates
 ($dy^2=dr^2+r^2d\phi^2$)  for $d=2+1$
  and in cylindrical coordinates ($dy^2=dr^2+r^2d\phi^2 + dy_3^2$) for $d=3+1$.
  The equations of motion for  the Ansatz (\ref{ansatz}) are given by
\bea z^{d-1}\partial_z\left(\frac{f}{z^{d-1}} \partial_z \psi\right)
+ \frac{1}{r}\partial_r(r\partial_r \psi) + \left (\frac{A_0^2}{f}
-\frac{(A_{\phi}-n)^2}{r^2} \right) \psi &=& 0\, , \nonumber \\
z^{d-3}\partial_z\left(\frac{f}{z^{d-3}} \partial_z A_{\phi}\right)+r \partial_r\left(\frac{1}{r} \partial_r A_\phi\right) - \frac{2 \, (A_\phi-n)}{z^2} \, \psi^2 &=& 0\, ,\nonumber\\
z^{d-3}\partial_z \left(\frac{\partial_zA_0}{z^{d-3}}\right)+
\frac{1}{rf} \, \partial_r\left( r \partial_r A_0\right)- \frac{2 \,
A_0}{z^2 \, f} \, \psi^2 &=& 0\, . \label{eom} \eea
We will impose regularity to our solutions. This requires at
$z=z_h$: \bea -\frac{d}{z_h} \partial_z \psi +
\frac{1}{r}\partial_r(r\partial_r \psi) -\frac{(A_{\phi}-n)^2}{r^2}
\psi &=& 0\, ,
\nonumber\\
-\frac{d}{z_h}\partial_z A_{\phi}+r \partial_r\left(\frac{1}{r} \partial_r A_\phi\right) - \frac{2 \, (A_\phi-n)}{z_h^2} \, \psi^2 &=& 0\, ,\nonumber\\
A_0&=&0\, , \label{bch} \eea while at $r=0$ we  must have
\bea
 \partial_r A_0&=&0\ ,\ \  A_{\phi} =0\,  ,\nonumber\\
  \partial_r \psi&=&0 \   \ \text{for}\ n=0\  , \  \   \psi=0 \ \ \text{for}\ n\neq 0\, .
  \label{bcr0}
  \eea
For a  superfluid, as we explained before, the energy of the
vortices is sensitive to the size of the sample $R$. We will
therefore limit $r\leq R$,  where $R$, as we commented before, is
taken much bigger than the vortex radius.

\subsection{Holographic superfluid vortices}\label{HolographicSuperfluid}

For a vortex superfluid configuration   $a_\phi$  is fixed:
\be
a_\phi=A_\phi|_{z=0}=\frac{1}{2}Br^2\,  ,
\label{bcsf}
\ee
where the constant $B$  represents  the external  rotation
(or, equivalently, the external magnetic field for a superconductor in a situation in which the magnetic field can be
considered frozen).
This corresponds to  a Dirichlet boundary condition at $z=0$.

Also we impose the following boundary conditions
at $r=R$:
\begin{equation}
 \partial_r\psi=0 \ , \quad \partial_r A_0=0\ , \quad A_{\phi}=\frac{1}{2}BR^2\, . \label{BCr=RSF}
\end{equation}
These conditions are consistent with the variational principle which is used to derive the equations of motion from the action.
The first two conditions represent the physical requirement that, far away from the vortex center, the solution should reduce to the superconducting/superfluid phase, which is independent of $r$, while the third one is a simple option compatible with (\ref{bcsf}).

\begin{figure}[h]
\centering
\begin{tabular}{cc}
\hspace{-0.1cm}
\includegraphics[scale=0.70]{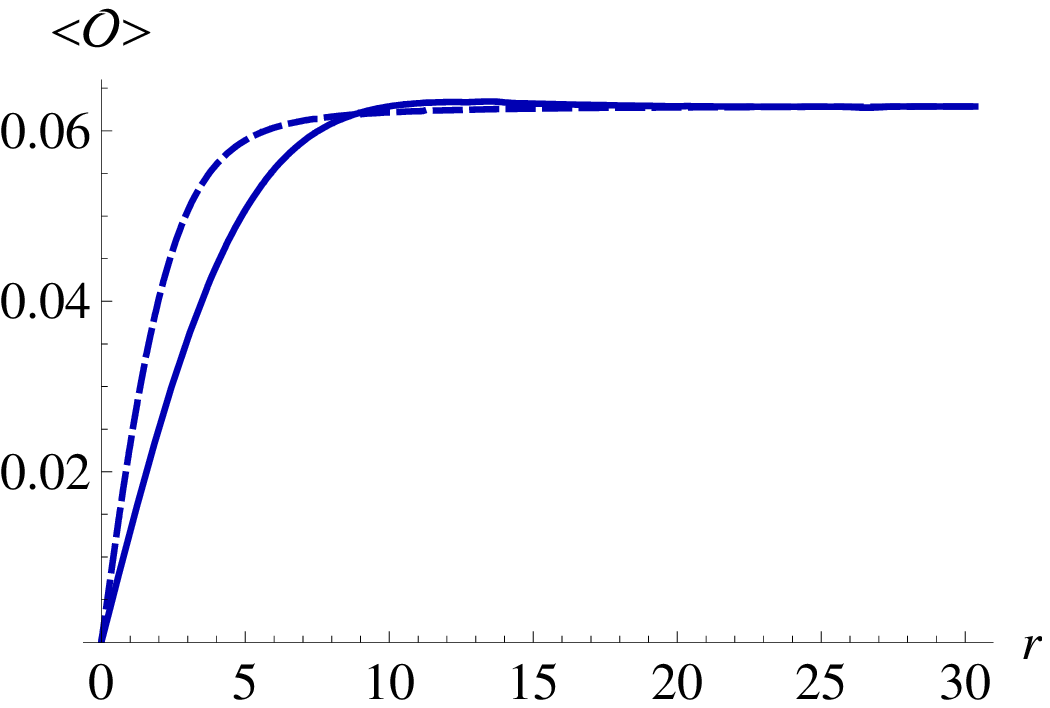} &
\includegraphics[scale=0.70]{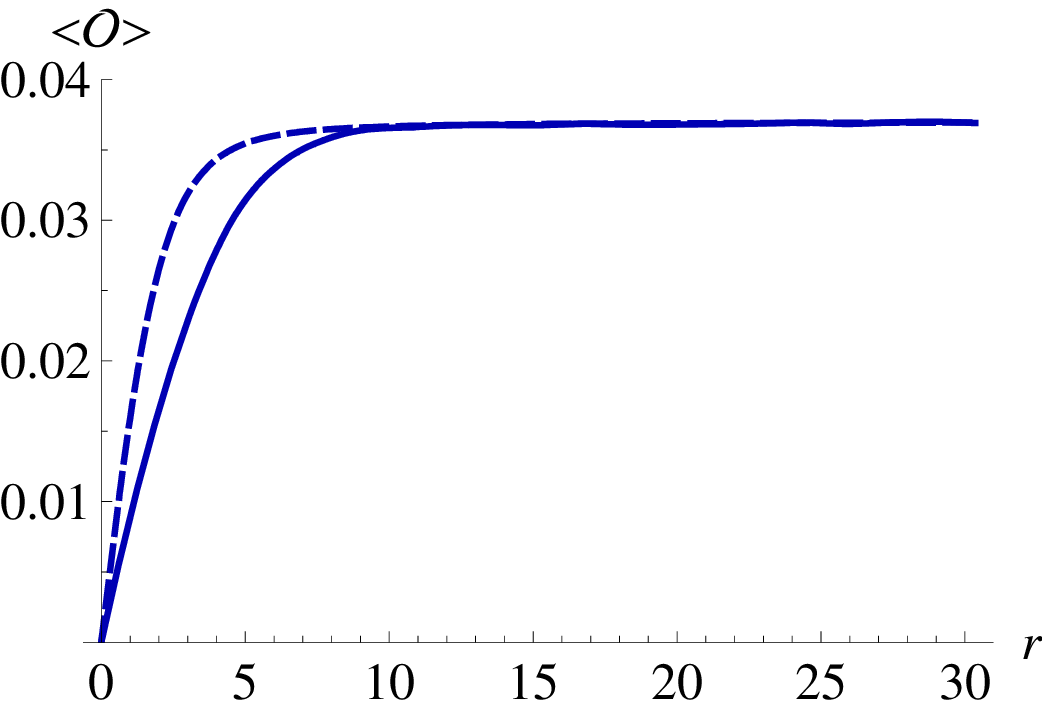}
  \\ \\
   \hspace{-0.1cm}
  \includegraphics[scale=0.70]{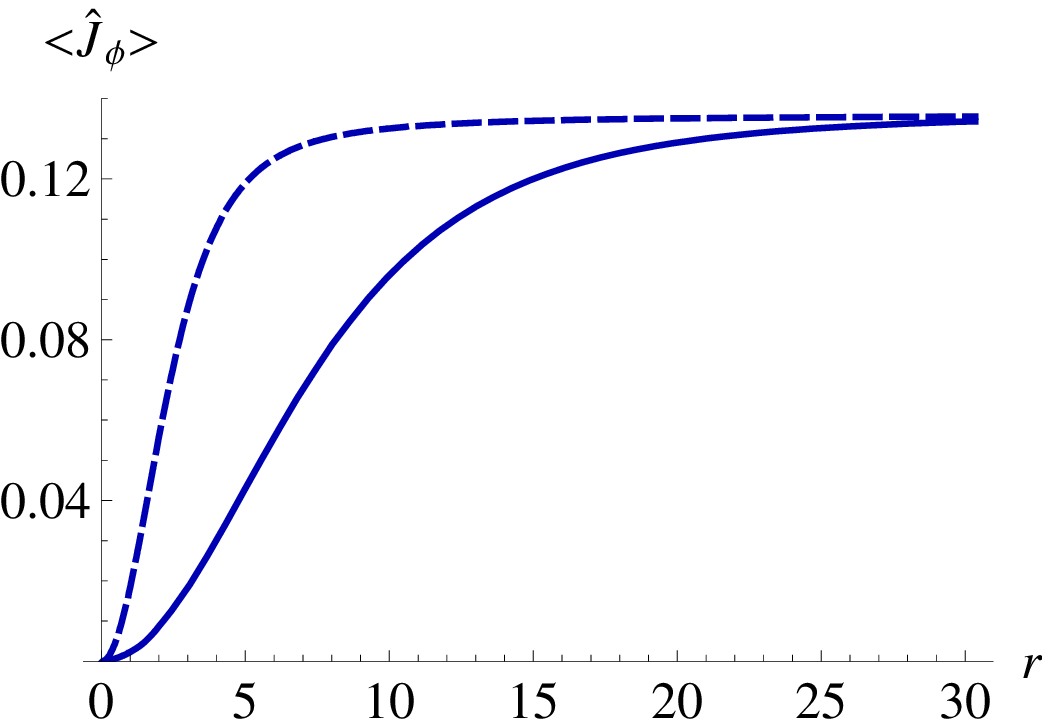}
   & \includegraphics[scale=0.70]{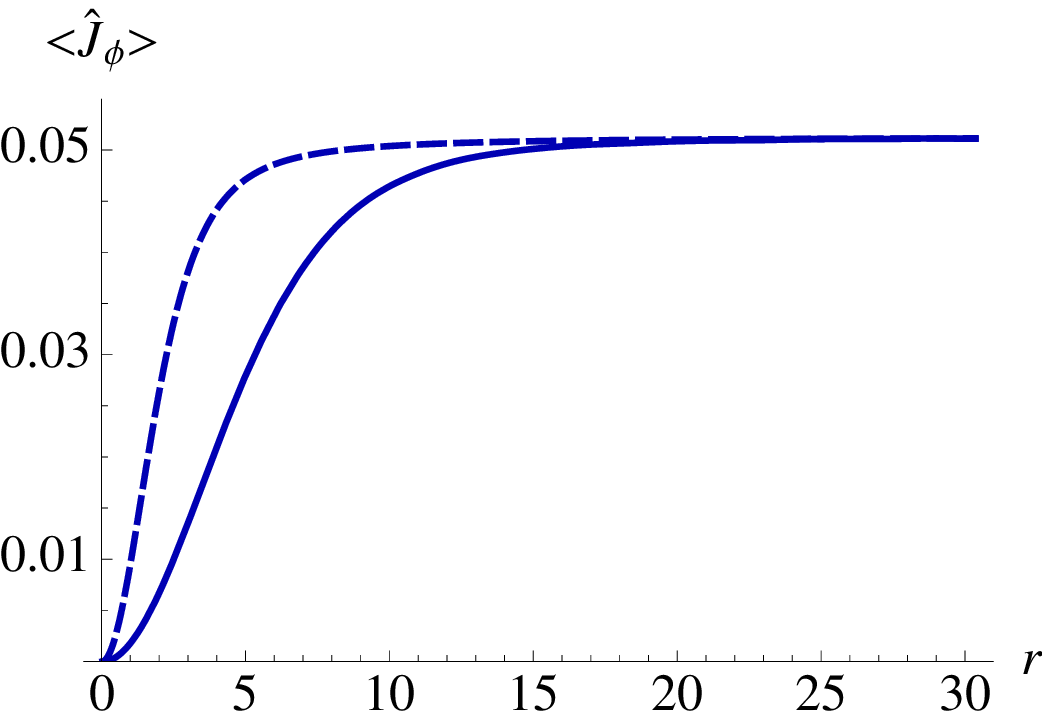}
\end{tabular}
\caption{\footnotesize The modulus of $\langle{\cal O}\rangle$ and $\langle
\hat{J}_{\phi} \rangle$ (up to a factor $L^{d-3}/g^2$) as functions of $r$ from the holographic
model in the $n=1$ superfluid  vortex solution for $d=2+1$ (solid
lines on the left) and $d=3+1$ (solid lines on the right).
In this plot we chose $T/T_c=0.3$ and  $B=0$.
The dashed lines are the corresponding  profiles in the GL model. Presented in units of $\mu=1$.}
\label{O}
\end{figure}

We have solved numerically Eqs.~(\ref{eom}) with the boundary conditions Eqs.~(\ref{bch}), (\ref{bcr0}), (\ref{bcboth}),  (\ref{bcsf}) and (\ref{BCr=RSF}) by using the COMSOL 3.4 package \cite{comsol}.
In Fig.~\ref{O} we present  the order parameter and the current as functions of $r$ for the $n=1$ vortex solution obtained from such numerical analysis.
Our solutions have the right behavior at $r\rightarrow 0$ and $r\rightarrow \infty$
as predicted in Eqs.~(\ref{rnbehaviour}) and (\ref{xibehaviour})  respectively.
We notice however that, unexpectedly,  the order parameter  $\langle{\cal O}\rangle$
develops a small bump at around $r\sim 12/\mu$, especially for the $d=2+1$ case.

It is interesting to know whether our results deviate from those of the simple GL theory.
For this purpose, we must first  specify the input parameters, $\xi_{\rm GL}$ and
$b_{\rm GL}$,  of the GL model.
We fit these two parameters from  two predictions of the holographic model: $B_{c2}$ and
$\langle \hat J_{\phi}\rangle$ at large $r$.
The value of $B_{c2}$ is  determined in the holographic model
as the value of $B$ at which $\langle {\cal O}\rangle$  reduces to zero everywhere in space.
With this value and   Eq.~(\ref{xiBc2})  we can obtain $\xi_{\rm GL}$.
From the value of $\langle \hat J_{\phi}\rangle$ at large $r$ as given in Fig.~1
we  match with the corresponding current in the GL model, which, for $B=0$, reads   $J_{\phi}^{\rm GL}(r\rightarrow \infty)=2n\left|\Phi_{\rm GL}(r\rightarrow\infty)\right|^2$; this allows to   obtain  $b_{\rm GL}$.
We find
\be
\xi_{\rm GL}\simeq 1.1\ (0.9)\mu^{-1}\ , \quad b_{\rm GL}\simeq 3.3\mu\  (12.4)\, ,
\label{valgl}
\ee
for $d=3\ (4)$ at  $T/T_c=0.3$.
Once $\xi_{\rm GL}$ and
$b_{\rm GL}$ are determined, we can obtain the prediction
 of the GL model for the condensate and the current
as functions of $r$ in the $n=1$ vortex configuration.
We show these  in Fig.~\ref{O}.
We can appreciate that  the holographic vortex differs significantly  from that of the GL theory. In particular,  the radius size of the vortex core in the holographic model is
considerably bigger than that
in the GL theory, namely $\xi> \xi_{\rm GL}$. For $T\simeq T_c$, however, the holographic model, like any other model of superfluidity, should reduce to the GL theory. We have checked numerically
that the holographic prediction for $\langle{\cal O}\rangle$ and $\langle \hat{J}_{\phi} \rangle$ approaches that of the GL theory when the temperature is very close to $T_c$.

Next, we calculate the vortex free energy $F_n$ of the $d$ dimensional superfluid.
It is then possible to verify that $B_{c1}$ behaves as predicted by the effective theory approach, Eq. (\ref{bc1}), and also that $F_{n}$ follows,
to a very good approximation, Eq.~(\ref{Fn2}).
Similarly, the results from the effective theory of Section~2, can explain the results obtained in Ref.~\cite{Montull:2009fe}.
Indeed, taking the value of $n_s(\psi_{\infty})=0.28 \sqrt{\rho}$  as  in Ref.~\cite{Montull:2009fe}, 
we obtain, from Eqs.~(\ref{mn2}) and  (\ref{Fn2}), 
\be
M_n\simeq 0.4  n R^2\sqrt{\rho}-0.1 R^4 \sqrt{\rho}  B\ ,\quad \frac{F_1-F_0}{\sqrt{\rho}}=0.9 \ln \left(R/\xi\right) -0.4R^2 B\, ,
\ee
that agrees \footnote{Here we point out a missprint in the value of $\beta_n$ given in Eq.~(24) of
Ref.~\cite{Montull:2009fe}: the correct one is $\beta_n\simeq 0.1(0.2)  R^4\sqrt{\rho}$.}, as well as
$B_{c1}$ in Eq. (\ref{bc1}),
with the numerical values obtained in  Ref.~\cite{Montull:2009fe}.

The value of $B_{c2}$ as a function of $T$,  that, as explained
before,  coincides with that of a superconductor ($B_{c2}=H_{c2}$),
will be presented in  Section~\ref{dynamicalholography}. As
discussed in Section~\ref{superconductor vortex}, any superfluid can
be considered as a deep Type II superconductor and therefore, when
$B$ is slightly smaller than $B_{c2}$, presents a triangular vortex
lattice. This property has been checked in Ref.~\cite{Maeda:2009vf}
for a holographic superfluid for $d=2+1$. Here we stress that the
same remains valid also for bigger values of $d$ as it uniquely
comes from the fact that, when the condensate is small, the theory
is well approximated by a GL theory.
 In the next section we will show that our holographic superconductor is a Type II superconductor and therefore is also
 characterized by a triangular lattice of vortices for $H$ slightly smaller than $H_{c2}$.


\subsection{Holographic superconductor vortices} \label{dynamicalholography}

To model an Abrikosov vortex we consider stationary configurations that do not possess a dynamical electric field but only a dynamical magnetic field.
Therefore at $z=0$ we will impose the boundary condition Eq.~(\ref{bcboth}) for $A_0$ and
Eq.~(\ref{maxwell2}) for $A_i$ that, in polar coordinates, reads
\be
\frac{L^{d-3}}{g^2}z^{3-d}\partial_z A_\phi\Big|_{z=0} +\frac{1}{e_b^2}r\partial_r\left(\frac{1}{r}\partial_r  A_\phi\right)\Big|_{z=0}=0\label{newbcsc}\, ,
\ee
where  we have taken $J^\mu_{ext}=0$.
At  $r=R\rightarrow\infty$ we impose that
\be
\partial_r\psi=0\ ,\ \ \  \partial_r A_0=0\ ,\  \ \  A_{\phi}=n
\label{rinftbc}\,.
\ee

\begin{figure}
\centering
\begin{tabular}{cc}
\hspace{-0.1cm}
\includegraphics[scale=0.70]{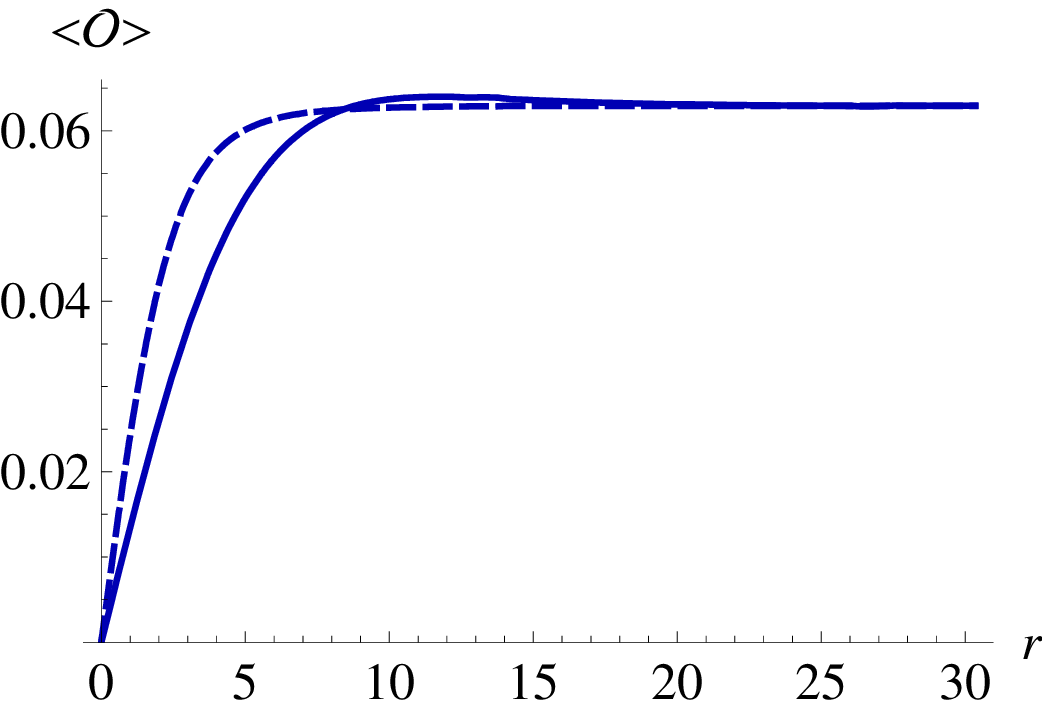}
& \includegraphics[scale=0.70]{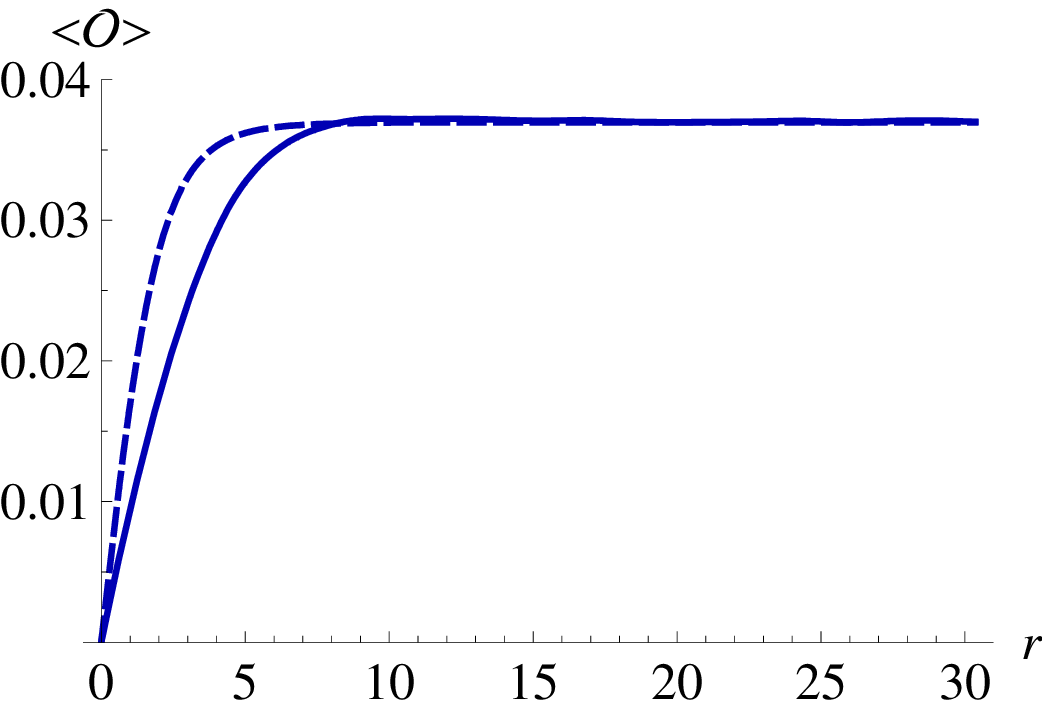}
 \\ \\
 \hspace{-0.1cm}
\includegraphics[scale=0.70]{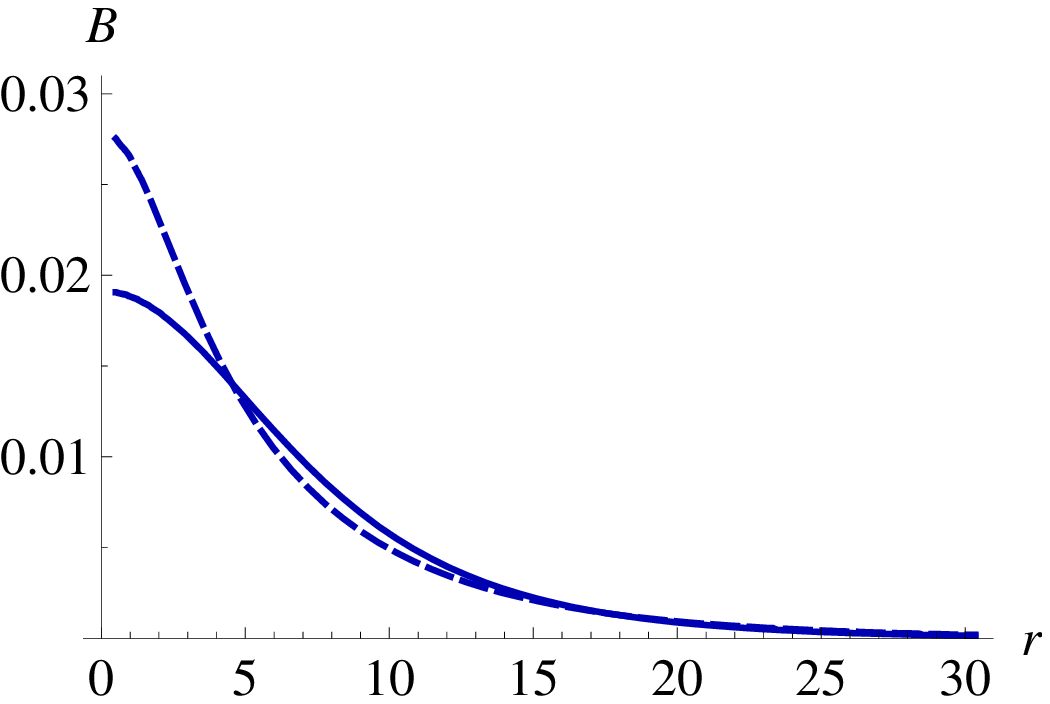}
 & \includegraphics[scale=0.70]{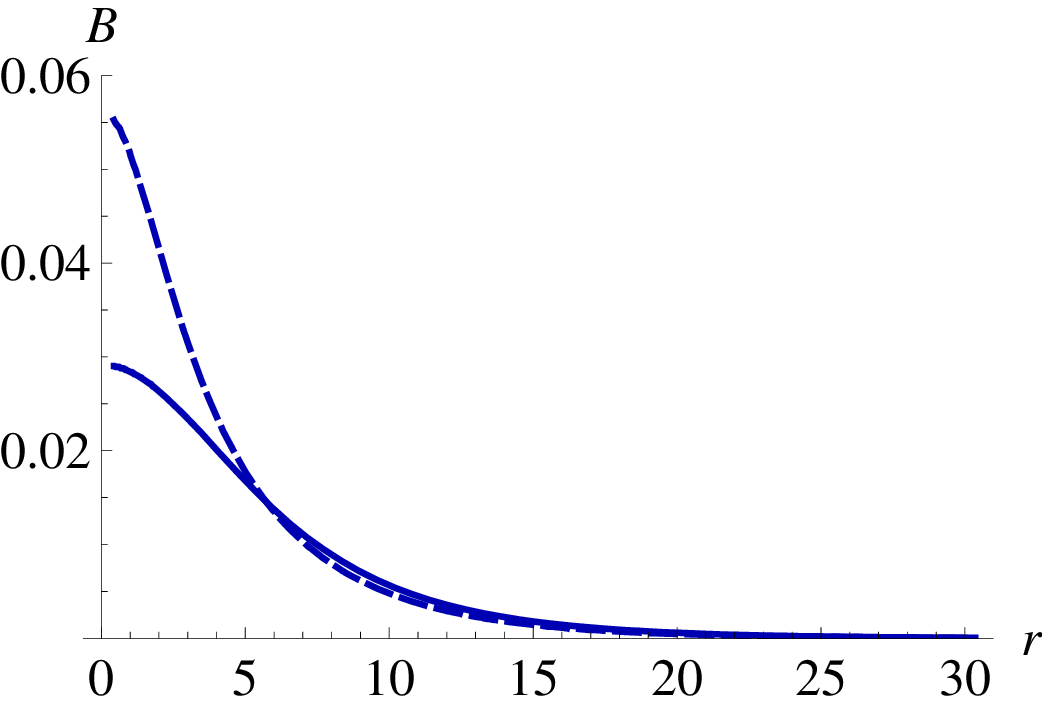}
\end{tabular}
\caption{\footnotesize The modulus of $\langle{\cal O}\rangle$ (up to a factor $L^{d-3}/g^2$) and $B$  as functions
of $r$ from our holographic model in the $n=1$ superconductor
vortex solution for $d=2+1$ (solid lines on the left) and $d=3+1$
(solid lines on the right). The dashed lines are the corresponding  profiles in the GL theory. Presented in units of $\mu=1$.}
\label{OSC}
\end{figure}

\begin{figure}
\centering
\begin{tabular}{cc}
\includegraphics[scale=0.70]{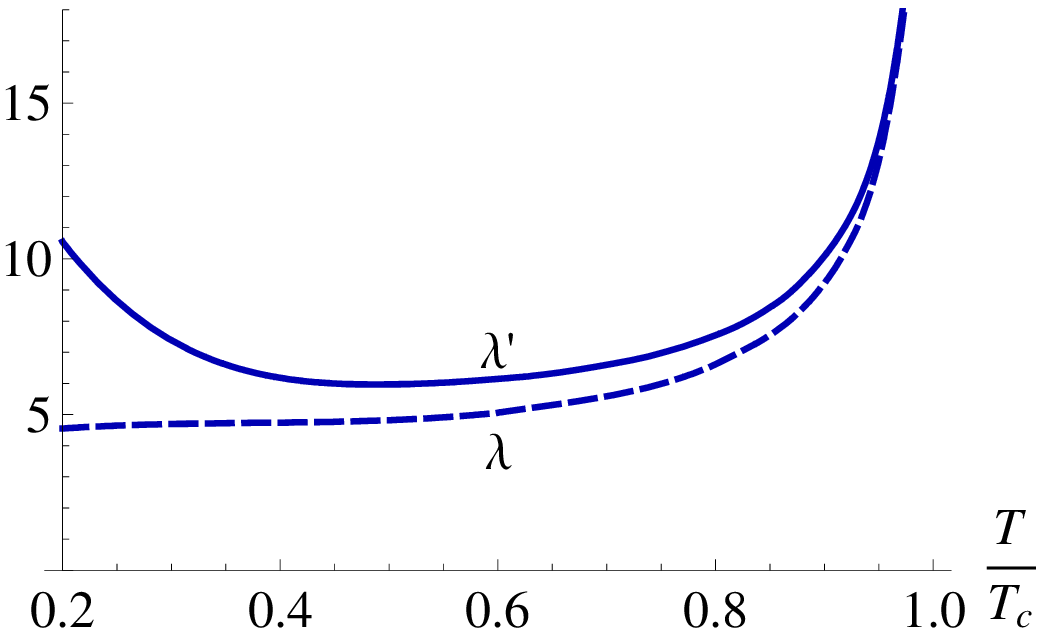} &
\includegraphics[scale=0.70]{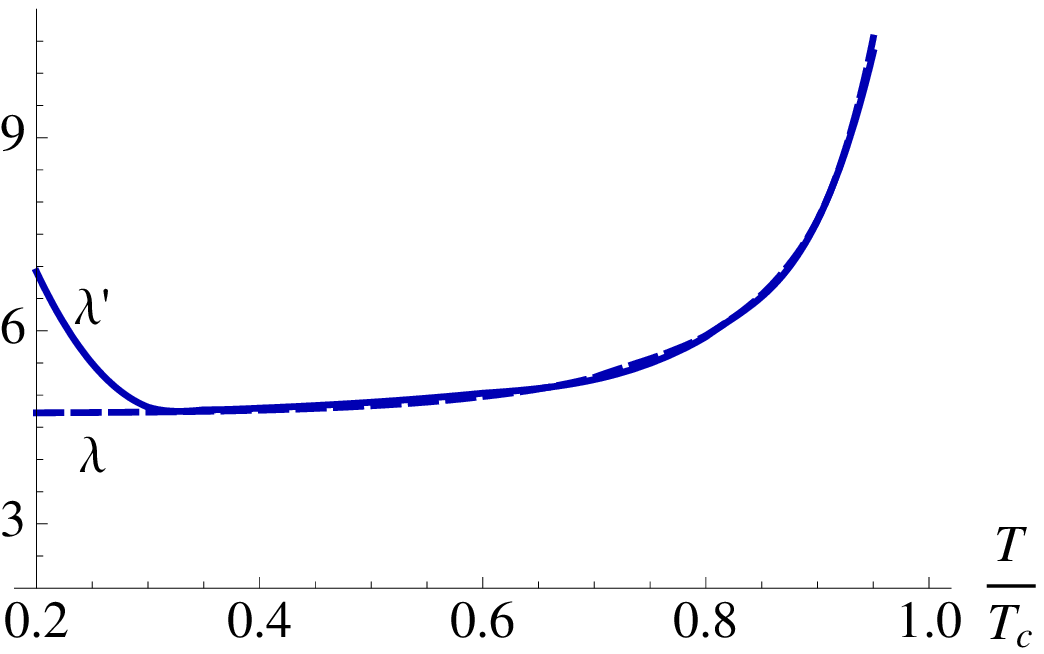}
 \end{tabular}
\caption{\footnotesize  $\lambda$ and
$\lambda'$ as functions
of $T$ from our holographic model for $d=2+1$ (on the left) and $d=3+1$
(on the right).  Presented in units of $\mu=1$.}
\label{lambda'}
\end{figure}

From the  set of equations (\ref{eom})
and boundary conditions
Eqs.~(\ref{bch}), (\ref{bcr0}), (\ref{bcboth}),  (\ref{newbcsc}) and (\ref{rinftbc}),
we can  numerically obtain the superconductor vortex configurations.
The profile for the condensate $ \langle{\cal O}\rangle$ and the magnetic field $B(r)=\partial_r A_\phi|_{z=0}/r$ are given as  functions of $r$   in Fig.~\ref{OSC}.
We have chosen $T/T_c=0.3$ and  $e_b/g\rightarrow \infty$ for $d=2+1$, while, for $d=3+1$,
we have taken  $e_b$ to satisfy $e_0^{-2}(T=T_c)\simeq 1.7 L/g^2$.
We observed that the fields have the expected  behavior at small and large $r$ given
by Eqs.~(\ref{limitr0}) and (\ref{limitrinfty}) respectively.
Indeed, the vortex profile at large $r$ has changed from the behavior of Eq.~(\ref{xibehaviour}) to that of Eq.~(\ref{limitrinfty}) as expected in an Abrikosov vortex with dynamical EM fields.
Similar to the superfluid case, however, the order parameter $\langle{\cal O}\rangle$
shows an unexpected slight  increase  at  around $r\sim 12/\mu$.
 In Fig.~\ref{lambda'} we show $\lambda$ and $\lambda'$,  defined respectively
 in   Eqs.~(\ref{lambda}) and (\ref{limitrinfty}), as functions of the temperature.
 For  $T\rightarrow T_c$ both quantities diverge  as expected, since in this limit we have $\psi_{\infty} \rightarrow 0$  and therefore  $\lambda'\rightarrow \lambda\rightarrow\infty$.
As $T\rightarrow 0$, however,  we observe that  $\lambda$ and $\lambda'$
differ considerably, with  $\lambda'$  increasing  its value at  $T/T_c\simeq 0.3-0.4$.
A priori, this would indicate that the magnetic flux tube becomes broader  as $T$ goes to zero, since the  penetration length $\lambda'$ grows.
Nevertheless,  we find that the situation is more complex; as $T\rightarrow 0$ the
  magnetic flux  develops two cores, one of size $\sim 1/\mu$  while the other $\sim \lambda'$.
This unexpected behavior deserves further studies.

 In Fig.~\ref{OSC} we also provide the corresponding curves in the GL theory; the parameters $\xi_{\rm GL}$ and $b_{\rm GL}$ in the
GL potential are fixed as in the superfluid case, Eq.~(\ref{valgl}), while the electric charge $e_0$ appearing in the GL action is determined by using the second definition in Eq.~(\ref{lambda}) applied to
the  GL case, that is  $\lambda_{\rm GL}=1/(\sqrt{2}e_0 |\Phi_{\rm GL}(r\rightarrow\infty)|)$,
 and by requiring $\lambda_{\rm GL}$ to be equal to  the value of $\lambda'$ of the holographic superconductor.
Again, as in the superfluid case, we  observe  that the radius size
of the vortex core is bigger in the holographic model than in the GL
theory. As expected,  we find these differences disappear as
$T\rightarrow T_c$.

\begin{figure}[h]
\centering
\begin{tabular}{cc}
\includegraphics[scale=0.70]{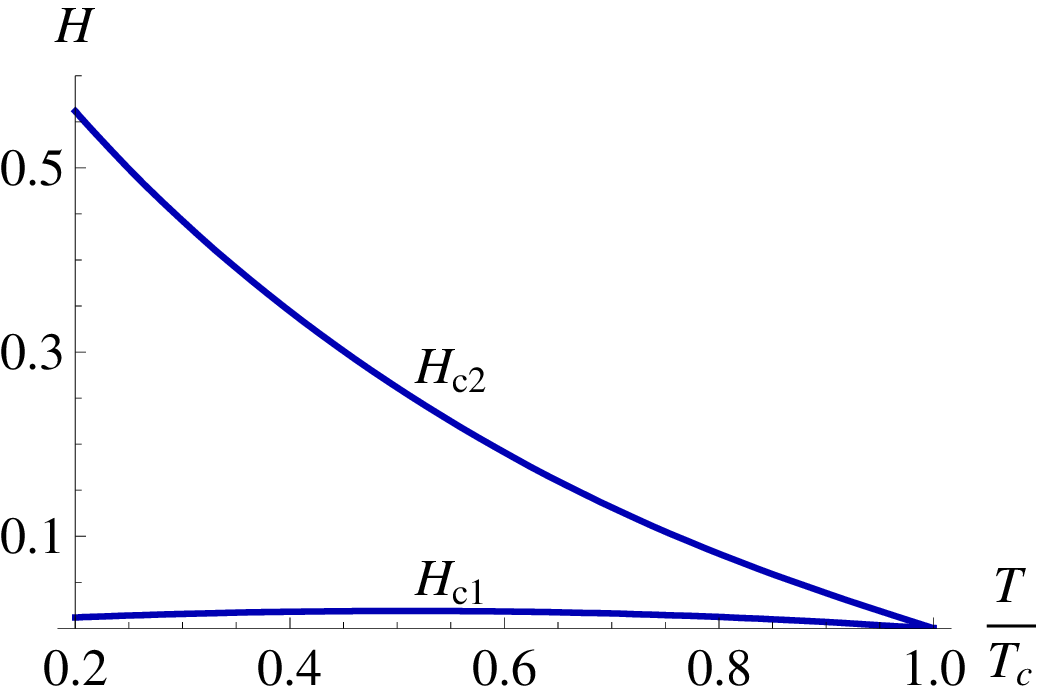} &
\includegraphics[scale=0.70]{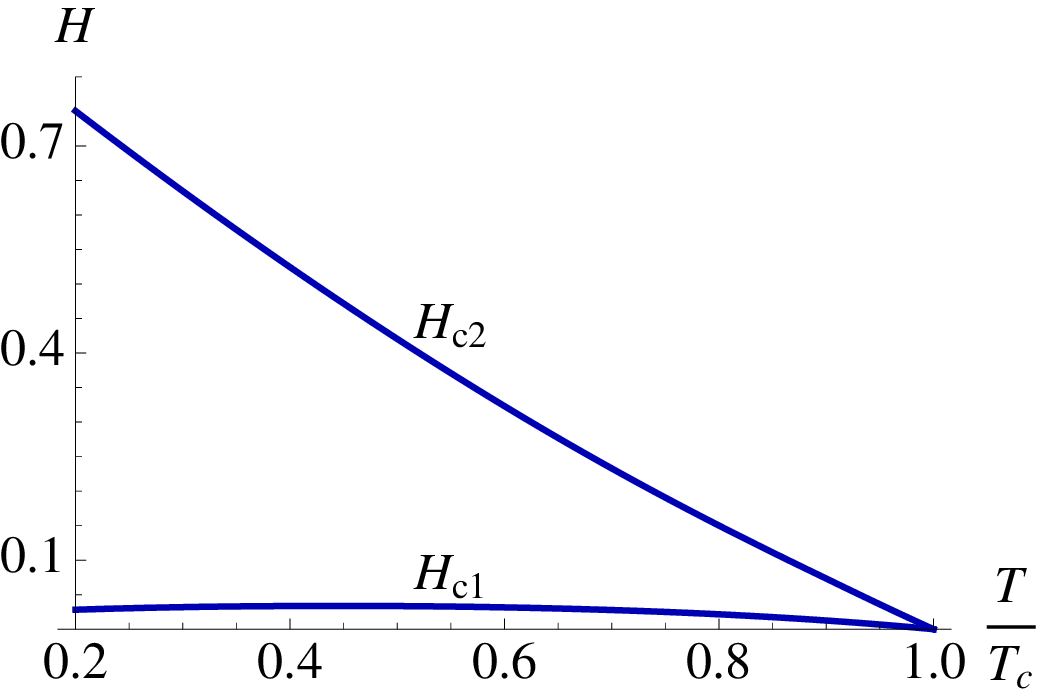}
\\ \\ \includegraphics[scale=0.70]{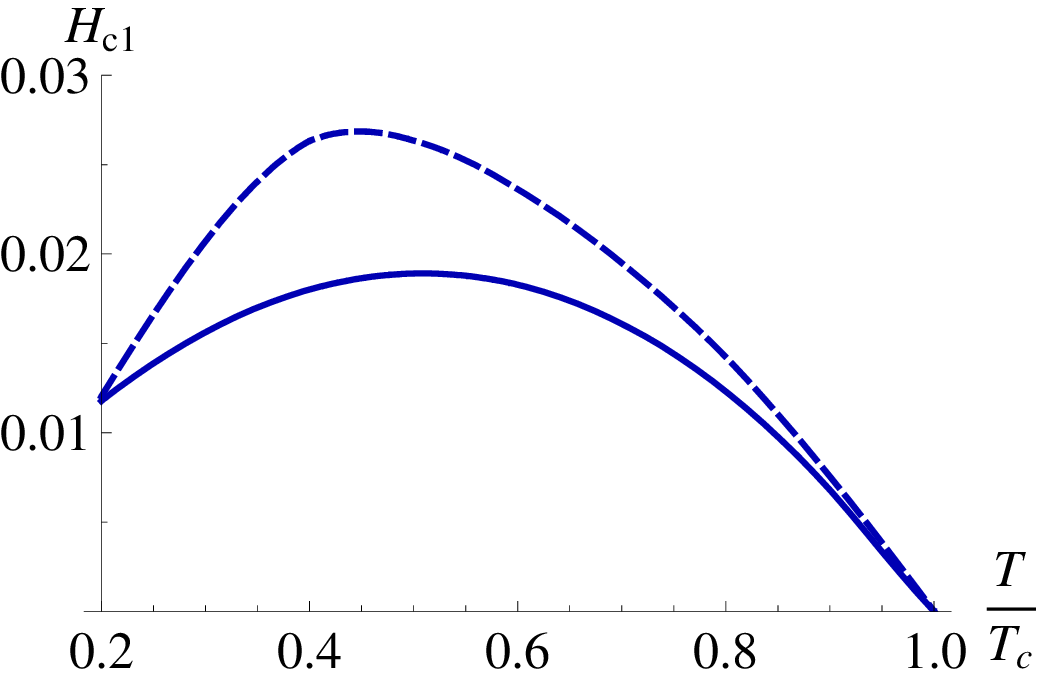} &
\includegraphics[scale=0.70]{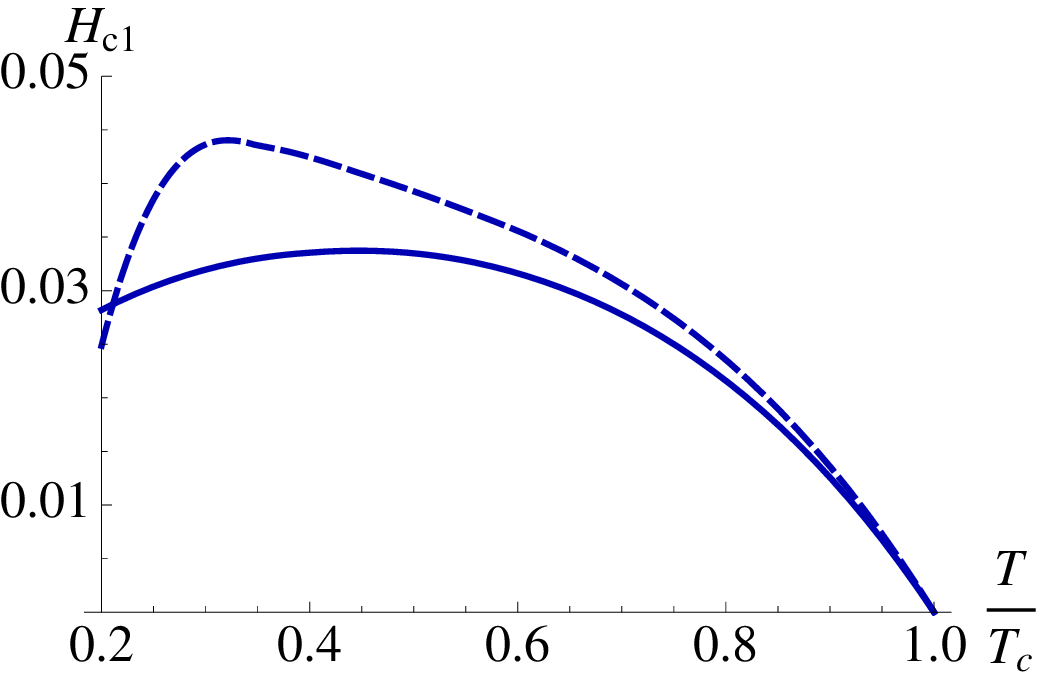}
\end{tabular}
\caption{\footnotesize $H_{c1}$ and $H_{c2}$ as functions of $T$
from our holographic model for $d=2+1$ (solid lines on the left) and
$d=3+1$ (solid lines on the right).  The dashed lines are
the corresponding  predictions for $H_{c1}$ from the GL theory.
Presented in units of $\mu=1$.} \label{Hc12}
\end{figure}

The free energy per unit of volume $V^{d-3}$ of the vortex configuration
 is, after taking into account the  kinetic term in  Eq.~(\ref{newg}),
  given by
\be
F_n=\frac{T}{V^{d-3}} S_E +2\pi  \int  dr r\frac{1}{2e_b^2}\frac{(\partial_ra_\phi)^2}{r^2}\, ,
\label{fnsc2}
\ee
where $S_E$ is calculated with the appropriated boundary conditions already stated.
Contrary to the superfluid case, we have checked that $F_1-F_0$ is  finite for $R\rightarrow\infty$
thanks to the presence of the gauge field.

To calculate the critical magnetic field
$H_{c1}$ we follow Eq.~(\ref{hhc1}).
 We find that   $H_{c1}<H_{c2}$ for any real value of  $e_0$.
This implies that for these holographic superconductors
 there is always a  range of values of $H$  for which vortex solutions are energetically favorable;
the superconductors are  always of Type II. In Fig.~\ref{Hc12} we
show $H_{c1}$ and $H_{c2}$ as functions of the temperature for the same values of $e_b$
as in Fig.~\ref{OSC}. Notice that $H_{c1}$ approaches zero as
$T\rightarrow 0$. This is due to our  normalization of $H$ in Eq.~(\ref{maxH})
that makes $H_{c1}\propto e_0^2$, which goes to zero as $T\rightarrow 0$.
We can, however,  derive
 $H_{c2}/H_{c1}\rightarrow\infty$ as $T\rightarrow 0$
 independently of such normalization. This is a  generic prediction of superconducting CFT.
Finally, we compare our results with those arising from  the GL
theory of superconductors. We observe that  $H_{c1}$ deviates from
the GL prediction for temperatures smaller than $T_c$.

 From the discussion given in Section~\ref{superconductor vortex},
 and the fact that our superconductors are of Type II,
 we know that the energetically favorable configuration when $H$ is slightly smaller than
 $H_{c2}$ is a triangular lattice of vortices.

\section{Conclusions}

We have shown how to introduce  a dynamical gauge field in holographic superconductors to study  vortex configurations.
In $d=2+1$  this  gauge field is part of  the CFT spectrum
and therefore can be considered to be an  emergent phenomenon, instead of an external field.
We have shown that vortex configurations, in the presence of this gauge field, follow  the  expected properties of   finite-energy Abrikosov vortices  where the magnetic field  drops exponentially
at  distances larger than  the vortex core.
We have calculated the energy of the vortices and   the critical magnetic fields
$H_{c1}$ and $H_{c2}$  that determine the  intermediate  (Shubnikov) phase.
In all cases we have found  that $H_{c1}<H_{c2}$
indicating that holographic superconductors are of Type~II.
For comparison,
we have also calculated the
vortex configurations in the absence of dynamical fields, corresponding to superfluid vortices,
and described their properties.

The vortex configurations found here differ considerably from those arising from a GL theory.
In particular, the vortex size comes out to be larger,  $H_{c1}$ has a different $T$ dependence, and, more importantly,
the penetration length  of the magnetic field differs significantly  as  $T\rightarrow 0$.

We have extended the study to
$d=3+1$ where a dynamical gauge field has to be introduced by a proper renormalization of the AdS-boundary terms. In this case, the gauge field does not respect conformal symmetry and  is  external to the CFT.
In spite of this, the vortex properties are found to be very similar to the $d=2+1$  case.

Although  we have focused on vortex solutions, the  method described here to introduce a dynamical gauge field  is general and can be used in other situations.
For example, one could  study the behavior of the EM field near the surface of a finite size superconductor
or in the junction between two superconducting samples in the presence of the Josephson effect \cite{Josephson:1962zz}.
Moreover, it would be interesting to extend the present analysis to 3+1 dimensional gauge fields sourced by a 2+1 dimensional CFT;
this would allow to study interactions between the EM field and layered superconductors.
Our studies can also be extended to $p$-wave holographic superconductors
\cite{pwave} and to holographic models that are dual to non-relativistic scale-invariant theories \cite{Balasubramanian:2008dm}.

\section*{Acknowledgements}
\label{sec:acknowledge}

We would like to thank A.~Sanchez for useful discussions. This
work was partly supported by CICYT-FEDER-FPA2008-01430, the ``Universitat Aut{\`o}noma de Barcelona'' PR-404-01-2/08, FPU grant AP2007-00420,
2009SGR894 and UniverseNet (MRTN-CT-2006-035863).

\end{document}